\documentstyle[psfig]{caosp}
\begin{document}

\pubyear{1998}
\volume{ }
\firstpage{1}

\title{Statistics on the spectral classification of CP2 stars in the
Southern Sky   } 

\author{M.~Rode, H.M.~Maitzen, E.~Paunzen} 

\institute{Institut f\"ur Astronomie der Universit\"at Wien,
T\"urkenschanzstr. 17, A-1180 Wien,
Austria (surname@astro.ast.univie.ac.at) }

\date {\today}

\maketitle

\begin{abstract}
A number of about 1500 spectroscopically classified CP2 stars in the
southern sky ($\delta =-90\degr$ to $-12\degr$) was extracted from the
Michigan Catalogue (Vols. I\,-\,IV).

This sample was compared with the classification from
Bidelman \& Mc Connell (1973).
We confirmed the spectral classification with the known
photometric peculiarity indices in the Geneva system and in the 3-filter
$\Delta a$ system (Maitzen 1976).
10\% of these stars show discordance between their respective 
types from the Michigan and Bidelman catalogues.
Several objects were measured with a CCD in the $\Delta a$ system in
spring 1995.
Eight stars are peculiar in $\Delta a$.
Furthermore, we have investigated the galactic distribution of all
programme stars. We conclude that the distribution resembles the one
of early-type stars, where the hotter (=\,Silicon) stars are more
concentrated towards the galactic plane than the cooler (=\,Strontium) 
objects.
\end{abstract}

\section{ Search and statistics for classified CP2 stars } 
About 30.000 stars with B, A and early F type classification were
extracted from the Michigan catalogue of two dimensional spectral types 
for the HD stars (Nancy Houk et al., 1975-1988; hereafter MC).
In the southern sky region from $\delta_{1900} = - 90\degr$ to
$-12\degr$, 1500 objects are classified as CP2 stars.
The limiting magnitude of this survey is 11, the maximum of the distribution
of magnitudes is between 8.5 and 10 for both subgroups.

74\% of all extracted CP2 stars are members of the Si-subgroup.
They are more concentrated towards the galactical plane than the cooler stars.
In the galactic area from $b = -15\degr$  to $b=+15\degr$, there are 84\% of
all hotter silicon stars and 58\% of all SrCrEu stars.
As expected, this result is in accordance with the galactic distribution of
main sequence stars.
A tendency to clustering for the investigated stars was not found.

Bidelman \& MacConnell (1973; hereafter BC) list 781 CP2 objects.
Their classification is based on the same objective-prism plates as the
sample of the Michigan catalogue.
Most of the compared stars have the same or a similar spectral
classification in MC and BC.

Photometric values are given for 981 stars in the Geneva system
(Rufener 1988) as well as in the 3-filter $\Delta a$ system
(Maitzen \& Vogt 1983).

430 out of 645 BC stars with known Geneva photometric indices, have a
significant peculiarity.

75\% of 339 BC-stars are peculiar in $\Delta a$. Three quarters of this
sample are members of the Si subgroup.
 
HD 134185 was classified as F0/2 V in MC (F2 V in the Geneva catalogue), but
 as Si in BC.
Si may be a typographical error meaning Sr, but the photometric $\Delta a$
 value confirms non-peculiarity.

The spectral class of HD 91756 is given as Fm $\delta $ Del in MC, (A0)  in
 the Geneva catalogue and Ap(SrCr) in BC.
The Geneva peculiarity index $\Delta (V1-G)$ shows no
 significant value for this star. 
 
 HD 110072 is identified as Ap(SrCr) in MC
and BC but as K0 in the Geneva Catalogue and the Simbad database.
The photometric values in the Geneva catalogue point to a late
A-type star.

A total of 78 stars are identified as non-CP2 stars in MC but as Si, Sr or
SrCrEu in BC. The spectral classification and photometric values 
in the Geneva catalogue are also different for these objects.
Therefore, we started an observing run with $\Delta a$ photometry
for a sample of them.

\section{ The $\Delta a$ measurements} 
 
Only 29 of the discrepant stars are peculiar in the Geneva photometric
system. 21 objects with known Geneva indices were measured with a CCD in
the $\Delta a$ system by H.M.~Maitzen and E.~Paunzen (Maitzen et al. 1997).

Observations were performed with the 61-cm Bochum telescope at ESO-La Silla on 
three nights from 29 to 31 May 1995. 
The three filters g$_{1}$, g$_{2}$ and y were used; their characteristics 
are listed in the following table.

\begin{center} 
\begin{tabular}{cccc}
\hline
\noalign{\smallskip}
\multicolumn{1}{c}{Filter}& \multicolumn{1}{c}{$\lambda_c$[\AA]} &
\multicolumn{1}{c}{FWHM\,[\AA]} & \multicolumn{1}{c}{transmission} \\
\noalign{\smallskip}
\hline
\noalign{\smallskip}
g$_{1}$ & 5027 &  222 & 66\,\% \\
\noalign{\smallskip}
g$_{2}$ & 5205 &  107 & 50\,\%   \\
\noalign{\smallskip}
y     & 5509 &  120 & 54\,\%   \\
\noalign{\smallskip}
\hline
\end{tabular}
\end{center}

$$ a = g_{2}-(g_{1}+y)/2 $$
$$\Delta a = a_{0}- a$$

We regard a star as photometrically peculiar if $\Delta a \ge 0.018$ or
$\Delta a \le -0.018$\,mag. To check the peculiarity found this way, the
Geneva criterion $\Delta (V1-G) > 0.010$\,mag
(North \& Cramer 1981) may be applied. We were unable to find here a 
significant influence of interstellar reddening on the $a$-values, which might
be expected from the relatively blue wavelength of $g_{2}$ compared to
those of $g_{1}$ and $y$. Obviously the rather low amount of reddening did
not succeed in producing sensible differential effects on the a-values.

Eight candidate stars (HD 84629, HD 86170, HD 128075, HD 129460,
HD 142778, HD 142960 and HD 167444) are to be regarded
as photometrically peculiar.
 
Comparing the classification performances of both sources (MC and BC) with
the help of the new available photometric evidence, one notes that for
eleven B-type stars with a giant luminosity class according to MC, the 
BC-peculiarity assignments are in better agreement with our results then 
the MC types.
Only in three cases are the Geneva-based $\log g$ values (North \& Kroll 1989)
typical of giants while the $\Delta a$ index shows no peculiarity.

\acknowledgements
Use was made of the Simbad database, operated at CDS, Strasbourg, France.

\end{document}